\documentclass[conference]{IEEEtran}

\usepackage{fmtcount}
\usepackage{graphicx}
\usepackage{courier}

\usepackage[dvipdfm, colorlinks=true, citecolor=Red, urlcolor=blue, pdfhighlight =/O]{}

\usepackage[usenames,dvipsnames]{pstricks}
\usepackage{epsfig}
\usepackage{pst-grad} % For gradients
\usepackage{pst-plot} % For axes

\begin{document}
%
% paper title
% can use linebreaks \\ within to get better formatting as desired
\title{IEEE 802.15.4 transceiver for the 868/915 MHz band using Software Defined Radio}

% author names and affiliations
% use a multiple column layout for up to three different
% affiliations

\author{
\IEEEauthorblockN{
Rafik Zitouni$^{\ddagger}$ $^{\dagger}$, Stefan Ataman$^{\ddagger}$, Marie Mathian$^{\ddagger}$ and Laurent George$^{\ddagger}$ $^{\dagger}$
}
\IEEEauthorblockA{
$^{\ddagger}$ ECE Paris-LACSC Laboratory \\
37 Quai de Grenelle, 75015, Paris, France\\
$^{\dagger}$ LISSI / UPEC \\
120, rue Paul Armangot\\
94400 Vitry S/Seine, France\\
Email: \{zitouni, ataman, mathian, lgeorge\}@ece.fr\\}
}

% use for special paper notices
%\IEEEspecialpapernotice{(Invited Paper)}

% make the title area
\maketitle

\begin{abstract}
%\boldmath
This paper reports an implementation of the PHY specifications of
the IEEE 802.15.4 standard for the frequency band 868/915 MHz on a
Software Defined Radio (SDR) platform. This standard is defined
for low power, low data rate and low cost wireless networks. These
specifications are used by the Zigbee technology for various
applications such as home automation, industry monitoring or
medical surveillance. Several hardware PHY 868/915 MHz band IEEE
802.15.4 transceiver implementations have been already reported on
ASIC and FPGA \cite{SabaterGL10} \cite{etal2006}. SDR offers one
possibility to realize a transceiver with high flexibility and
reconfigurability~\cite{Ulversoy10}. The whole --transmitter and
receiver-- chain has been defined in software using the GNU Radio
software project \cite{Blossom2004} and the USRP (Universal
Software Radio Peripheral) platform from Ettus Research
\cite{Ettus}. Two new blocks have been added to the GNU Radio
project, one for the Direct Sequence Spread Spectrum and the
second for the reconstruction of the packets. The experimentations
have been performed in a noisy environment and the PER, BER and
SNR have been computed. The obtained results are coherent with
what can be expected from the theory.
\end{abstract}


\section{Introduction}
% no \IEEEPARstart
Most of the standards and protocols of lower layers of wireless
transmissions (AM, FM, IEEE 802.11, IEEE 802.15.1, IEEE 802.15.4.
etc.) are mainly implemented in hardware (HW). This lack of
reconfigurability makes the adaptation to varying radio resources
difficult, especially when multiple standards need to be often
switched in order to take advantage of the scarce radio resources
available. The purpose of Software Defined Radio (SDR) is to avoid
these drawbacks of traditional wireless communications and replace
the hardware equipment by software. The huge advantage of SDR
platform lies in its flexibility, its multi-functionality and its low
development cost. The reconfigurability of the platform ensures
the reusability of the hardware~\cite{Ulversoy10}, thus minimizing
the design complexity of new RF terminals.

%The performance of the SDR can be improved by the use of FPGAs
%(Field-Programmable Gate Arrays) for parallel processing of
%multiple applications, thus reducing its size, cost and time of
%development. The SDR technology has the potential to replace
%complex equipment such as: BSS (Base Station Subsystem), receiver
%stations for the satellite communications, GPS (Global Positioning
%System) receivers, military transceivers. Furthermore, adding a
%layer of cognitive radio algorithms allows the adaptation of the
%SDR platform to the available context by dynamically allocating
%its radio resources.

%The evolution of the SDR is the result of the several projects, in
%particular governmental development programs,  \emph{e.g.} the
%SpeakEasy I, the SpeakEasyII~\cite{Pucker2003} and the ongoing
%Joint Tactical Radio System (JTRS)~\cite{North2006} programme, the
%Terminal Radio Software (TERSO) programme~\cite{SpanishSDR}. We
%can cite also the French ``Software Radio Architecture'' Plan
%d'\'Etude Amont (PEA)~\cite{GNSV07},  and the European Secured
%Software Defined Radio Referential (ESSOR)\cite{ESSOR} programme.

The ideal SDR allows the analog-to-digital (ADC) and
digital-to-analog (DAC) conversion to be as close as possible to
the antenna \cite{mistola93}, eliminating the need of
high-frequency radio subsystems. Subsequently, the CPU executes
the software (SW) subsystem of the SDR, all signal processing operations are accomplished by SW. Unfortunately, today's technology is neither cost-effective for direct ADC conversion from the
antenna nor enough power full to compute GSPS (Giga
Samples-per-Second) in real-time. Therefore, the typical SDR
platform available today uses HW high-frequency radio front-end, the SDR part being
implemented in the baseband only. The HW supporting the SDR
platform is typically based on FPGAs or DSPs (Digital Signal
Processors)~\cite{Sadiku2004}.

%Several commercial and open source SDR platform are available
%today, having varying characteristics. Examples in commercial SDR
%platforms are Microsoft's SORA (featuring a frequency band from
%2400MHz to 5000MHz and a sampling rate up to 20MSPS
%\cite{Tan_2009}) and Typhoon SDR (frequency band from 200MHz to
%4000 MHz with fully adjustable transmit/receive sample rates from
%2.4 Ksps to 100 Msps\cite{Typhoon}).

GNU Radio~\cite{GNURadio} and OSSIE~\cite{OSSIE} (Open-Source
Software Communication Architecture Implementation Embedded) are the two
open source software subsystems for the USRP (Universal Software
Radio Peripheral) SDR from Ettus Research \cite{Ettus}. The USRP
HW is available in different versions. In our implementation we
used the USRP1 HW, featuring a sampling rate of 128 MSPS (Mega
Samples-per-Second) for the transmitter and 64 MSPS for the
receiver. By addition of different daughter-boards, the
baseband signal can be transposed in frequency bands up to 6000
MHz. The USRP1 HW platform proves to be also cost-effective,
compared to its competitors (Microsoft's SORA and Datasoft's
Typhoon).

%(priced at 700\$, versus 5000\$) for

The IEEE 802.15.4~\cite{2011a} standard defines the physical and link layers
for low-rate Wireless Personnel Area Networks (LR-WPAN), used in
wireless sensor networks applications with strong energy
consumption constraints. The physical layer comprises three
principal frequency bands allowing 49 channels: 16 channels in the
2450 MHz for the ISM (Industrial Scientific Medical) band, 30 for
North America and 3 channels in the 868 MHz band for Europe
\cite{2011a}. The band of 2450 MHz operates at law data rates of
250 kb/s while the bands of 915 MHz and 868 MHz operate at 40 kb/s
and 20 kb/s respectively. 

%The 2011 specifications of IEEE 802.15.4
%\cite{2011a} allows the use of different modulation techniques and
%data rates for the specified channels. The D-BPSK (Differential
%Binary Phase Shift Keying) is one of these modulation techniques
%used in the 915/868 MHz band. These frequencies are of interest
%because they allow to communicate at longer ranges.

A number of hardware implementations of the IEEE 802.15.4 have
been reported on ASICs or FPGAs \cite{SabaterGL10},
\cite{etal2006}, but they do not allow us to control the flexibility and the
ability of all software stack layers. The first software
implementation of the IEEE 802.15.4 using the GNURadio environment
for the 2450 MHz band was reported in \cite{Schmid2006}. In wireless sensor networks, the transceiver in the 868/915 MHz band is more suitable when low data rate transmission are used between sensor nodes. Furthermore it presents a longer range than that of
the 2450 MHz band for a given link budget. The objective of our work is to implement the specifications of the IEEE 802.15.4 standard for the 868/915 MHz band, which is not yet reported in the literature. 

%The authors in \cite{Muller_2008} propose the implementation of
%the Data Audio Broadcast transceiver with OFDM (Orthogonal
%Frequency Division Multiplexing) modulation/demodulation. The IEEE
%802.11p standard encoder have been developed in \cite{fuxj10}, the
%objective is to simplify the complex OFDM processing. In
%\cite{GnuSounder}, the PN sequence based channel sounding system
%have been implemented and tested.

Our software transceiver was developed by closely following the
IEEE 802.15.4 specifications for the 868/915 MHz bands. The
implementation is similar to the one of 2450 MHz band presented
in \cite{SchmidDS06}, \cite{Schmid2006}. To evaluate the
transmitter/receiver performances, the BER (Bit Error Rate) and SNR (Signal-to-Noise Ratio) have been
computed by changing the input power signal at the transmitter.

The rest of the paper is organized as follows. Section II presents
a description of the SDR platform used. In Section III, we present
the description of the developed transmitter/receiver chain.
Section IV discusses the experimentations and the obtained
results. Finally, in Section V we formulate some concluding remarks.

%\hfill
%\hfill
\section{USRP and GNU Radio}
In the following two subsections we describe briefly the USRP1 HW
\cite{Ettus}, used in our implementation as well as the GNU Radio
\cite{Blossom2004} toolkit.

\subsection{Universal Software Radio Peripheral}
\begin{figure}[!h]
\begin{center}
\begin{tabular}{c}

\includegraphics[height=6cm,width=8cm]{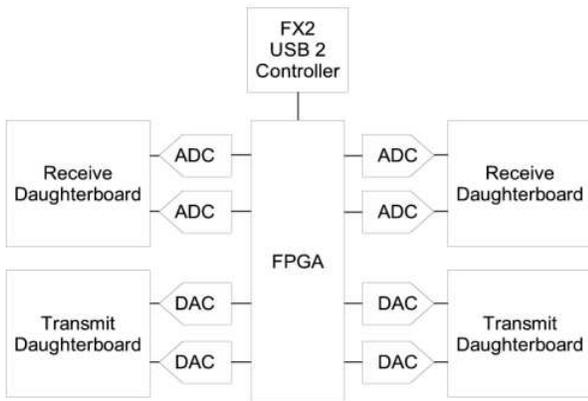}\\
\end{tabular}
\caption{USRP1 block diagram \cite{Ettus} } \label{fig-1}
\end{center}
\end{figure}

The USRP1 HW consists of a motherboard and optional add-on RF
daughterboards. It is connected to a host computer via USB 2.0.
The USRP's motherboard supports up to four daughterboards: two for
transmission (TX) and two for reception (RX).
%Each daughterboard is able to tune into a different range of
%frequencies from 0 to 6000MHz and connected to DAC/ADC and to an
%antenna via SMA interface.
The motherboard has four 12-bits ADCs (with a maximum sampling
rate of 64 MSPS), four 14-bit DACs (with a maximum conversion rate
of 128 MSPS), and an Altera FPGA for simple but high-speed
operations such as up-conversion, down-conversion, interpolation,
and decimation~\cite{Ettus}. The ADCs and DACs allow us to receive
baseband signals up to 32 MHz and are able to generate baseband
signals up to 50 MHz. Unfortunately, the USB tunnel limits these
performances to 8 MHz. The USRP1 provides buffer in both the USB
controller and the FPGA at 2 KB and 4 KB respectively.
Fig.~\ref{fig-1} depicts the USRP1 blocks from the motherboard.

\subsection{GNU Radio}
GNU Radio is an open source project toolkit for building software
radios that run on host computers \cite{Blossom2004}. It provides
signal processing blocks for modulation, demodulation, filtering
and various Input/Output operations. New blocks can be easily
added to the toolkit. The software radio platform is created by
connecting these blocks to form a \textit{flowgraph}. The blocks
are written in C++ and they are connected through a Python script.
The Verilog HDL layer is dedicated to configure the FPGA.

\begin{figure}[!h]
\begin{center}
\begin{tabular}{c}
\includegraphics[height=3.5cm,width=6.5cm]{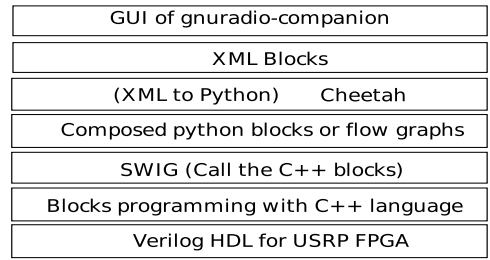}\\
\end{tabular}
\caption{Software layers of the GNU Radio } \label{fig-2}
\end{center}
\end{figure}

The advantage of Python in connecting these processing blocks is that
it allows the data flow to be at maximum rate, without being
interpreted. The integration of the C++ blocks into the scripting
language is provided by the SWIG (Simplified Wrapper and Interface
Generator), which is an interface compiler. Many signal processing
blocks are available to the GNU Radio community to facilitate the
development. To create a flow graph we can proceed by the
graphical interface called gnuradio-companion or directly through the
python code. The C++ blocks are described by the XML code to
facilitate the use and the visibility of the block chains, the XML
is interpreted to the python code by the cheetah
tools\footnote{http://www.cheetahtemplate.org/} . In Fig.
\ref{fig-2} we depict the programming language layers of the GNU
Radio.

\section{Transceiver Description}

\begin{figure*}[!t]
\begin{center}
\begin{tabular}{c}
\includegraphics[height=4.1cm,width=13cm]{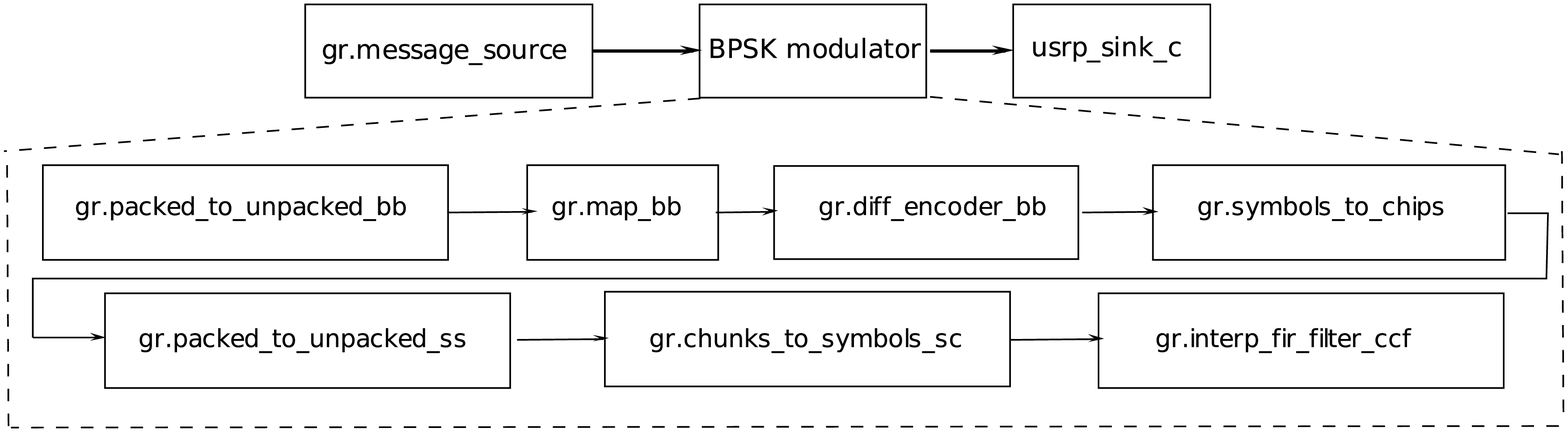}\\
\end{tabular}
\caption{Transmitter flow graph} \label{fig-3}
\end{center}
\end{figure*}

The IEEE 802.15.4 \cite{2011a} standard is the definition of
wireless physical (PHY) and medium access control (MAC) protocols
for low-data rate and low power applications. It specifies two
families of bands: the first one is centered at 868 and 915 MHz with
20 and 40 kbps, the second one at 2450 MHz with 250 kbps.

%In wireless sensor networks, the transceiver in the 868/915 MHz
%band is more suitable when low data rate transmission are used between
%sensor nodes. Furthermore it presents a longer range than that of
%the 2450 MHz band for a given link budget.

The specifications from \cite{2011a} define the use of different
modulation techniques and data rate for the specified channels.
The D-BPSK (Differential Binary Phase Shift Keying) is one of the modulation techniques used in the 915/868
MHz. The symbol spreading is the Direct Sequence Spread Spectrum
(DSSS), in which each symbol is represented by a Pseudo Noise
sequence of 15 chips. The chips are modulated/demodulated by the
D-BPSK encoding/decoding at rates of 300
kchips/s and 600 kchips/s for the 868 MHz and 915 MHz bands
respectively.

\subsection{Transmitter}
\begin{figure*}[!t]
\begin{center}
\begin{tabular}{c}
\includegraphics[height=1.3cm,width=10cm]{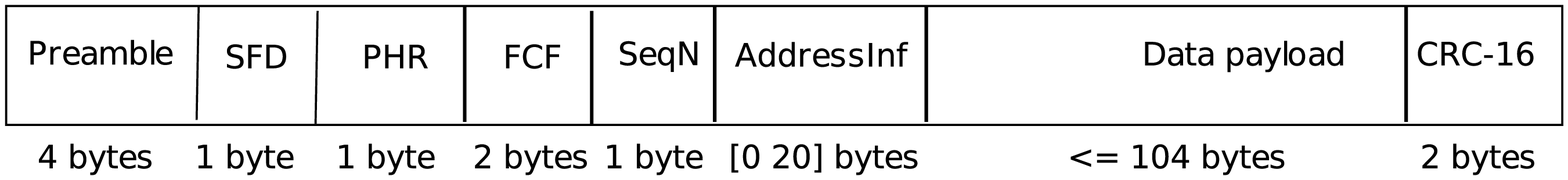}\\
\end{tabular}
\caption{IEEE 802.15.4 packet format for the USRP } \label{fig-4}
\end{center}
\end{figure*}

Our transmitter comprises eight processing blocks, as depicted in
Fig. \ref{fig-3}. The definition of the packet messages is based
on that of the IEEE 802.15.4 standard. The packet format
is detailed in Fig. \ref{fig-4}.  At the output of the
transmitter, the maximum packet size is $133$ bytes. Due to the
USB 2 tunnel, the packet size should be a multiple of 128 samples,
therefore, zero padding with the x/00 (representing the
``\textit{NUL}'' character) is performed. The number of padded
bytes is conditioned by the parameter called $Byte\_Modulus$ which
depends on the sampling rate and on the number of bits per symbol.
The $Byte\_Modulus$ is given by:
%If a packet size proposed by the user  modulo the $Byte\_Modulus$
%is zero, we don't padding, otherwise, the modulo result represent
%the bytes number add at the end of the packet.
\begin{equation}
Byte\_Modulus = \textrm{LCM}\left(\frac{128\textrm{
MSPS}}{8\textrm{ MSPS}},
sps\right)\cdot\left(\frac{bps}{sps}\right)
\end{equation}
where
\begin{itemize}
 \item $128$ MSPS -- DAC sampling rate of the USRP1 % ---
 \item $8$ MSPS -- Sampling rate of the USB tunnel % ---
 \item $sps$ -- Number of samples per symbol %---
 \item $bps$ -- Number of bits per symbol %---
 \item $\textrm{LCM}$ -- Lowest Common Multiple of $16$ MSPS and
 $sps$
\end{itemize}
To avoid padding and to get the same fields as in the IEEE
802.15.4 specifications, the packet size is set equal to 130
bytes. This size is obtained by reducing the address information
field $AddressInf$. Moreover, a 16-bit CRC (Cyclic Redundancy Check) is attached to the
packet payload, allowing the receiver to calculate the PER (Packet
Error Rate).

\begin{table}[!h]
\begin{center}
\begin{tabular}{|p{1.5cm}|p{4cm}|}
\hline
\scriptsize{{\bf Input bits}} & \scriptsize{{\bf Chip values} (c0 c1 $\dots$ c14) } \\
\hline
\scriptsize{0} & \scriptsize{1 1 1 1 0 1 0 1 1 0 0 1 0 0 0}  \\
\hline
\scriptsize{1} & \scriptsize{0 0 0 0 1 0 1 0 0 1 1 0 1 1 1}  \\
\hline
\end{tabular}
\caption{Symbol to chip mapping}
\label{tab-1}
\end{center}
\end{table}

The packets are divided into chunks of symbols by the
\texttt{gr.packed\_to\_unpacked} block, each symbol representing $1$ bit.
Since the C++ programming language does not allow us to have a data type of
1 bit, the bits in the bytes of an input stream are grouped into
chunks of 1 byte. The MSB (Most Significant Bit) of 8 output bits
represents the one bit at the input of \texttt{gr.map\_bb}. After that, the differential
encoder \texttt{gr.diff\_encoder\_bb} encodes a current symbol modulo-2 of the previous one. Then,
the symbols are mapped by \texttt{gr.symbols\_to\_chips} into 15 Pseudo Number Sequence chip as
specified in Table \ref{tab-1}. The output of mapping is short-type
(16 bits carrying the 15 chips). With the same technique the
stream is unpacked to a chunks of 16 bits representing the chips
stream. Each chip is represented by a complex constellation point
in 1 dimension for the BPSK modulator by \texttt{gr.chunks\_to\_symbols\_sc}. The stream is then fed through
a Root Raised Cosine \texttt{gr.interp\_fir\_filter\_ccf} filter which up-samples the signal,
after which it is sent from the host computer via USB to the
transmitting USRP.

\subsection{Receiver}

The receiver begins with an USRP source connected to a squelch
filter \texttt{gr.pwd\_squelch} which admits only signals with a certain dB
strength. The squelch filter in GNU Radio outputs 0 when the
incoming signal is too weak. The stream result of the squelch is
passed to the Automatic Gain Control \texttt{gr.agc\_cc} (AGC) of a D-BPSK
demodulator, it regulates the gain in a way that does not have a
large or small amplitude and to avoid distortions. After that, the
result enters to two filters in \texttt{gr.interp\_fir\_filter\_ccf}, FIR (Finite Impulse Response) and RRC
(Root Raised Cosine) allowing the receiver to process the change
of the transmitted pulse and minimize symbol interference. The RRC
filter makes the correlation between the received signal and the
expected one. It calculates a FIR filter coefficient or a
tap weight. The demodulator synchronizer is composed by two
blocks, a Costas Loop \texttt{gr.costas\_loop\_cc} (Phase Locked Loop) and the Mueller and
M\"uller \texttt{gr.clock\_recovery\_mm\_cc} \cite{Danesfahani1995}. The Costas Loop recovers the
carrier and improves the Bit Error Rate of BPSK demodulator. The
Mueller-M\"uller Timing recovery block recovers the symbol timing
phase of the input signal. After the demodulator, the stream is
converted from complex to real in order to send it to our
developed block \texttt{ieee.ieee802\_15\_4\_packet\_sink} which slices real stream from
chips to bits. With the knowledge of the packet length field, the
packets are decoded. The first information decoded is the preamble
with four 0x00 bytes, it is followed by the rest of the fields. If
the preamble is not detected, the preamble search is re-launched.
The receiver performs the error detection without correction.
After the packet construction, a CRC-16 value is processed and compared to that carried by the CRC field of the received frame. If they
are not equal, the received packet is incorrect.

The packet queue is observed by an external
python thread. When a message arrives to the queue, a thread
starts to call a function that process the packet, e.g: like printing
the packet content.

\begin{figure*}[!t]
\begin{center}
\begin{tabular}{c}
\includegraphics[height=3cm,width=17cm]{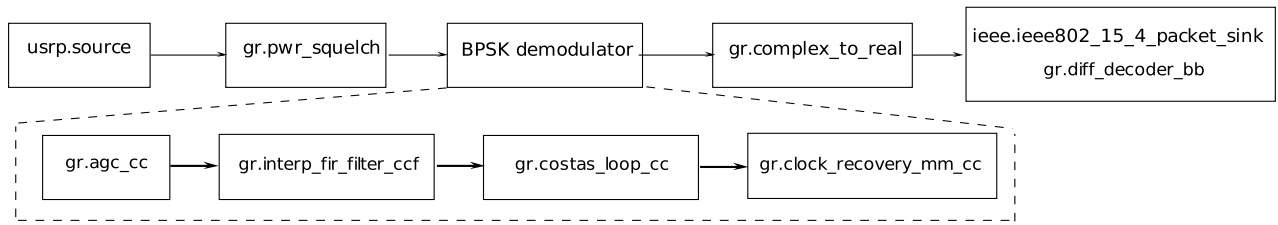}\\
\end{tabular}
\caption{Receiver flow graph} \label{fig-5}
\end{center}
\end{figure*}

\section{Experimental conditions and results}

The experimentations are performed in an indoor environment. We
use two USRP1 platforms coupled with RFX 900 daughterboards covering a frequency range from 750 MHz to 1050 MHz. The GNU Radio
software stack is executed on a host computer having one Core 2
Duo CPU running at 2.4 GHz and 2 GB of RAM. The distance between
the two USRP1 boxes was greater than 2 meters.

% --- COMMENT: REVIEW THIS PART
% --- START

The principal USRP1 parameters are the transmitter Interpolation
$I$ and receiver Decimation $D$, they are calculated according to
a symbol rate $r$, $DAC\_s$ and $ADC\_s$ sampling, and a number of
samples per symbol $sps$, such as:
\begin{equation}
I=\frac{DAC\_s}{r\cdot sps},\qquad D=\frac{ADC\_s}{r\cdot sps}
\end{equation}
where :
\begin{itemize}
\item $DAC\_s = 128$ MSPS
\item $I\in [16,\:20,\:24,...\:508,\:512]$
\item $ADC\_s=64$ MSPS
\item ${D\in [8,\:10,\:12,...\:254,\:256]}$
\end{itemize}

For 20 kbps, the transmitter and receiver parameters  are
respectively $I = 400$ and $D = 200$ with a $sps = 16$. Otherwise,
when the data bit rate is equal to 40 kbps, the $I$ and $D$ take
the same values but with $sps = 8$. The amplifier amplitude is
defined by a dimensionless scalar with values ranging from $0$ to
$32767$.

\begin{figure}[!h]
\begin{center}
\begin{tabular}{c}
\includegraphics[height=6.1cm,width=8.2cm]{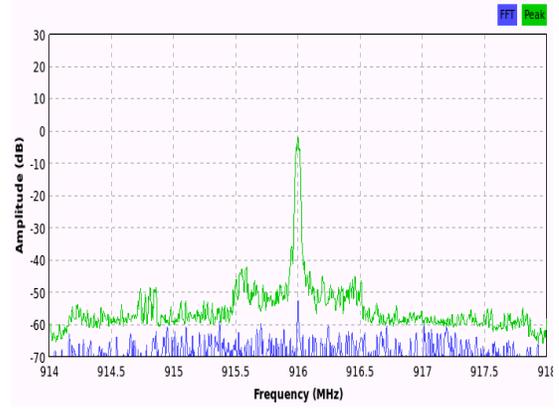}\\
\end{tabular}
\caption{Power spectrum of our software transceiver recorded with the USRP and drawn by FFT gnuradio plot } \label{fig-6}
\end{center}
\end{figure}

The results shown in Fig. \ref{fig-6} depict the power spectrum of
the transmitted signal from the GNURadio transmitter. They
correspond to the output of the FFT spectrum-analyser tool that is
included in the GNURadio framework. A peak is visible with our
software transceiver when we choosing the channel at 916 MHz, with a number
of $35$ samples per symbol which allow us to have an intermediate
frequency of $1.5$ MHz. This value is in concordance with the
values taken by the transmitted power spectral density of the IEEE 802.15.4 standard
(see Fig. \ref{fig-6}). Furthermore, frequencies at the edge of
the main band are visible but strongly attenuated. These
imperfections may be due to the roll-off characteristics of the
interpolation filter in the up-conversion processing of the FPGA.

% --- STOP

\begin{figure}[!h]
\begin{center}
\begin{tabular}{c}
\includegraphics[height=6cm,width=8cm]{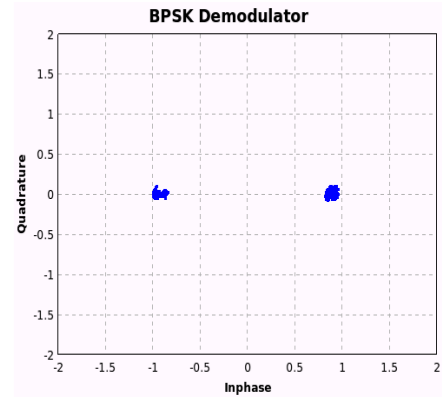}\\
\end{tabular}
\caption{Receiver symbol constellation } \label{fig-7}
\end{center}
\end{figure}

We use a D-BPSK modulation and the receiver constellation  is
depicted in Fig. \ref{fig-7}.

\begin{figure}[!h]
\begin{center}
\begin{tabular}{c}
\includegraphics[height=6.1cm,width=7.2cm]{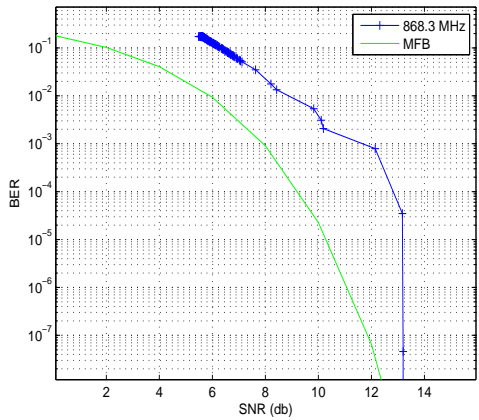}\\
\end{tabular}
\caption{The BER versus received SNR for central frequency 868.3 MHz and for the MFB} \label{fig-8}
\end{center}
\end{figure}

The performance of D-BPSK modulation is evaluated without packet
generation. The flexibility of the GNU radio permits the
reconfigurability of the transmitter/receiver chain by adding or
replacing blocks. In a first experiment, we use the modulator and
demodulator chains to measure the BER and SNR parameters. Fig.
\ref{fig-8} illustrates the average BER versus the input SNR (dB)
for the frequency $868.3$ MHz and for the MFB Matched Filter Bound of D-BPSK modulation. The results have been computed by changing the amplifier amplitude values from 1000 to 12000 with the step of 100 for a time period of $400$ seconds. Although noisy, the results are in concordance to the theory, proving that the implementation is working.

\begin{figure}[!h]
\begin{center}
%\begin{tabular}{c}
\includegraphics[height=6.2cm,width=7.2cm]{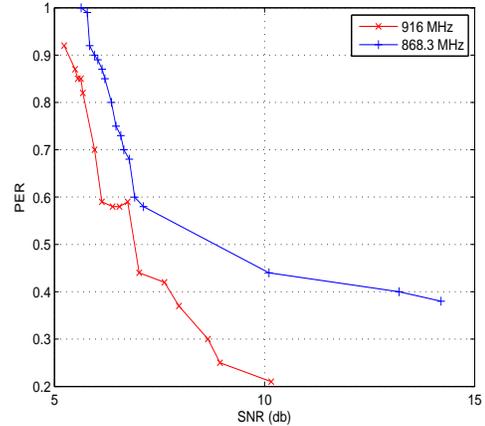}\\
%\end{tabular}
\caption{The PER over received SNR using two central frequencies 916 MHz and 868 MHz} \label{fig-9}
\end{center}
\end{figure}

In the second experiment, the packet generator and packet sink are
connected to the transmission chain and we measure a PER parameter as a function of SNR (dB). The measures are
obtained by sending for each amplifier amplitude 100 packets apart
from 0.2 s between two successive packets. The PER decreases when
the amplifier amplitude increases. The shape of the curve is
compliant to that of the BER (show Fig.\ref{fig-9}). The PER depends on the synchronization between the transmitter and the receiver. We noticed that the synchronization does not
occur at every execution. This issue may arise when the
USRP does not clear its buffer memory.

%The ratio of the packets transmitted and received with a correct
%CRC is 80\%.

\section{Conclusion}
In this paper, we report the implementation of the IEEE 802.15.4
standard on a SDR transceiver for the 915/868 MHz band. The SW
stack is based on the GNURadio open-source project and the HW is
based on an USRP1 platform from Ettus Research. The BER and PER of
the 802.15.4 have been calculated independently in an indoor
environment by changing the signal amplitude. The results are
coherent with the lower theoretical bound that is expected. The
obtained performances of the PER are degraded compared to the BER
because the successful receiving packets depend on the
synchronization and the BER.

%
%\section*{Acknowledgment}
%
%
%The authors would like to thank...

\bibliographystyle{IEEEtran}
\bibliography{Biblio}

% that's all folks
\end{document}